\documentclass[a4paper,11pt]{article}
\usepackage{pos}
\usepackage[capitalise,noabbrev]{cleveref}
\usepackage{natbib}

\title{2’ science: A Science Communication Project for Astrophysics}
\ShortTitle{2'science}
\author[a]{Eirini Batziou}
\author[b]{Stella Boula}
\author[c]{Marios Kalomenopoulos}
\author[d]{Dimitris Kantzas}
\author[e]{Apostolos Spanakis-Misirlis}
\author[f]{Vasileios Spanakis-Misirlis}

\affiliation[a]{Max Planck Institute for Astrophysics, Karl-Schwarzschild-Str. 1, 85748, Garching, Germany}
\affiliation[b]{Department of Physics, National and Kapodistrian University of Athens, Panepistimiopolis, GR 15783 Zografos, Greece}
\affiliation[c]{Institute for Astronomy, University of Edinburgh, Royal Observatory, Edinburgh EH9 3HJ, UK}
\affiliation[d]{Anton Pannekoek Institute for Astronomy (API) \& GRavitational AstroParticle Physics Amsterdam (GRAPPA), University of Amsterdam, Science Park 904, 1098 XH Amsterdam, the Netherlands}
\affiliation[e]{Department of Informatics, University of Piraeus, Greece}
\affiliation[f]{School of Rural and Surveying Engineering, National Technical University of Athens, Greece}

\newcommand{\ts}{2'science}

\emailAdd{2sciencegr@gmail.com}

\abstract{
Two-minute science (\ts) is a science communication project initiated and supported by early-career Greek astrophysicists. With this endeavor, which started in December 2020, we try to bridge the gap between the scientific community and the public. This project is based on the simple idea of writing short articles with an approximate reading time of two minutes. These articles cover several topics and their difficulty scales to cover a broad audience range, from young students to experienced adults. We support the idea of “ask an expert” in Astrophysics in Greece, where any reader can pose a question. We offer the appropriate answer either by writing it ourselves or by contacting the field experts from the Greek astronomical society. Furthermore, our previous science communication experience leads us to design educational activities for students and/or adults based on pedagogical means. A successful one was an “escape-zoom” titled “Escape to Other Worlds”, a digital version of an escape room. Further activities are Astronomy workshops for teenagers, online talks to schools, and our participation in a scientific podcast to trigger the public interest in Astrophysics. We communicate this work through social media, where several thousands of people already follow our work.
}
\FullConference{37$^{\rm{th}}$ International Cosmic Ray Conference (ICRC 2021)\\
		July 12th -- 23rd, 2021\\
		Online -- Berlin, Germany}

\begin{document}
\maketitle

\section{Introduction}
Gazing at the Universe has always been very attractive to humankind. Over the thousands of years on this planet, people have always been amazed by the night sky and have been looking to answer its mysteries. Whether on a scientific level or just due to their passion for Astronomy, people want to learn more and seek further knowledge. Our eyes to the Universe might be a binocular and/or a telescope but science has progressed a lot in various aspects making it difficult for the interested public, as well as scientists of other fields, to follow up.

One of the responsibilities of modern researchers is to bridge the gap between the scientific community and society by making contemporary knowledge accessible to everyone. In Greece, for instance, there are a few telescopes that operate as scientific facilities but are also open to the public during star gazing nights. That triggers the interest of people covering a broad range of ages, from very young kids to older people who did not have a chance to observe through a telescope for decades.

Passionate people about the Universe also seek out further material such as articles, books, scientific web pages to study and broaden their knowledge. Even if many people understand English, the scientific terminology might be incomprehensible to those who are not familiar with these expressions. Moreover, it is crucial for students to have access to scientific material written in their native language.

There are multiple fields in modern Astronomy that have been historically proven to be more popular compared to others. For instance, our solar system has always been attractive to the public because observing a planet that is light minutes away with a telescope might be something unique. Moreover, space technologies are quite popular especially since the very first human moon landing. 

However, there are multiple new sub-fields of Astronomy and Astrophysics that were developed and advanced just the past few decades. Some of these fields are: High-Energy Astrophysics, Astroparticle physics and others. Overall, people are not familiar with these newly born areas, and thus, it is the responsibility of the active researchers in these fields to communicate and popularise them.

\section{The project of \ts}
Driven by our passion for Astronomy, we decided to form a team of young researchers in December 2020 to communicate our passion to the public audience. The group members are located in different countries around Europe, with the majority of them living in Greece. The main idea is to write short but comprehensive articles with no more than 500 words, discussing some of the main topics of various fields of Astronomy, including High-Energy Astrophysics and Astroparticle physics. We store all of our material on a dedicated website\footnote{\href{https://sites.google.com/view/2sciencegr/home}{https://sites.google.com/view/2sciencegr/home}}. Such relatively small texts can be read in approximately two minutes; hence we decided to name the whole project \ts\ (two-minute science). 

\subsection{Aims}
The main goal of the \ts\ project is to bring the scientific community closer to the public audience and vice versa. We want to bridge the gap between scientists and the interested readers who are always in the lookout for the most recent advances in physics. We try to give them the chance to learn more about topics and fields that are not widely spread and hardly found in the Greek language, such as Radio Astronomy, High-Energy Astrophysics, Astroparticle physics, modern topics in Cosmology and others.

Furthermore, we request the assistance of other young researchers, even if they are not members of the \ts\ group, to introduce their field of expertise and answer relevant questions posed by the audience (see below). We thus give the floor to any member of the Hellenic Astronomical Society\footnote{\href{https://helas.gr}{https://helas.gr}}, but also invite scientists from all over the globe. This process allows all of us to improve our writing skills and develop new ones, such as editing and reviewing public articles.

\subsection{Who we are}
The team of \ts\ consists of a small group of young researchers and some auxiliary members who are mainly occupied with the visual and artistic side of the project. Our core team consists of 12 members at the moment, while we had more than 25 guests contributing in different ways. In the past, some of the \ts\ members have also been part of the outreach team of the National Observatory of Athens, responsible for guiding tours and educational activities.

\ts\ is a team-oriented project and forwards the team spirit rather than the one-person initiative. We thus try to learn from each other and combine the different ways of communicating our passion for Astrophysics. We also want to emphasize the significant contribution from all of our guests who play a crucial role in this project.

\subsection{Methods}
Within 2’science we utilize a variety of methods to make physics, and Astrophysics in particular, accessible to a broader audience. Every target group has different characteristics and hence our activities try to incorporate the necessary pedagogical elements for every age and its interests. These range from more “classical” methods, such as a website and public talks, to more creative initiatives that involve interactive activities for the public. We describe them in turn below.

Our website hosts a number of educational resources: 

- \textbf{Popular science articles about physics and Astrophysics, covering so far:} Astrophysics, Space Physics, Technology, and Miscellaneous topics. We want these articles to be accurate and concise, hence  more than one member of the team are responsible for the editing and in many cases, the assistance of a more senior researcher is requested. The articles are arranged into three levels of difficulty (+: easy, ++: moderate, +++: advanced) which correspond to the background required. 

- \textbf{“Ask an Expert” Questions and Answers section. }Inspired by the very successful work of websites like “Ask an Astronomer” \footnote{\href{http://curious.astro.cornell.edu}{http://curious.astro.cornell.edu}}, we have developed a dedicated section in which questions from the public are answered by experts in the relevant field. We follow the same thematic structure as in the articles. Many of the answers provide extra educational references for further reading.

- \textbf{Fun facts.} Designed for social media engagement, we are developing a database of interesting, brief facts about Astrophysics for the Greek community. Apart from the usual, exciting information, we also try to offer a physical description involved in each case. This differentiates them from the “standard” physics facts, that focus mainly on impressing the audience since we add an educational perspective that, in our opinion, is generally missing in similar concepts.

- \textbf{Public talks.} All the members of the team, either individually or as smaller groups, have contributed to many public talks of various levels of difficulty. Podcasts, school lectures, Universities’ Open days, talks at observatories' visitor centers, and amateur astronomical societies are some of the activities we have done so far.

Apart from the above, more mainstream techniques of science communication, we are always starving to develop new methods that can connect scientists and science with the wider public. Some of our most recent attempts are:

- \textbf{Escape Zoom:} During the pandemic, many of our activities had to be moved to an online format. In the past, a part of the team has organized a “Planet Marathon”. A treasure hunt, with checkpoints all over Athens, where the participants had to solve Astrophysics quizzes and riddles in order to advance. The success of this event, motivated us to try something similar in an online format. For that reason, we organized an “Escape Room” activity within Zoom\footnote{\href{https://zoom.us}{https://zoom.us}}, around the theme of “Exoplanets”. The event started with a brief, introductory lecture on “Habitability in Galaxies” and then the main activity followed: our world was hit by a GRB\footnote{This was a nice opportunity to demonstrate to the public the real power of a GRB and its realistic effects on our planet - see, for example, \href{https://iopscience.iop.org/article/10.1086/496914}{(B. Thomas et al. 2005)} and \href{https://journals.aps.org/prl/abstract/10.1103/PhysRevLett.113.231102}{(T. Piran and R. Jimenez, 2014)} - in the unfortunate scenario that its beam is directed towards Earth.} and the teams had to find a suitable exoplanet in order to save humanity. We had five main stations, each devoted to a specific exoplanet. Each team had to solve specific riddles connected to each station, including quizzes, simple physics experiments, and using software for detecting exoplanets among other things, in order to successfully pass the station and get the physical properties of its exoplanet. When all stations were completed successfully, the teams had to find which one of their exoplanets was the most suitable for terrestrial life, in order to win. We had 70 participants, in teams of 1 or more persons each, and 12 organizers helping at all stations. The feedback was very positive and we are planning to organize similar events in the future.

- \textbf{A Radio Telescope for All: PICTOR!} PICTOR\footnote{\href{https://pictortelescope.com}{https://pictortelescope.com}} is an open-source radio telescope that enables anyone to observe the sky in radio wavelengths, using its convenient web platform for free. The goal is to introduce students, educators, astronomers, and others to the sky at radio frequencies while promoting radio Astronomy education without the need of building a large and expensive instrument. PICTOR can be remotely operated by anyone at any time, and its ease of use does not require an in-depth STEM background by the observer.

The radio telescope is primarily designed and intended for observations of the 21\,cm hydrogen line, a spectral line commonly observed by radio astronomers to study the distribution of neutral hydrogen in the Universe. Observations with PICTOR allow users to measure the Doppler shift and radial velocity of Galactic clouds of neutral hydrogen, identify the spiral structure of the Milky Way and even map out patches of the sky in radio wavelengths. As of May 2021, the telescope hosts over 6,000 observations on the archive, conducted by more than 1,000 unique users from 50+ countries around the world, and has been extensively used by schools and universities for hands-on Astronomy experiments.

- \textbf{AstroLab for young Greek scouts}. We organized an online AstroLab with Greek scouts, with more than 30 participants (25 teenagers) and 6 organizers from our team. There we had demonstrations and introductory lectures about our Galaxy, scales in the Universe, and Radio Astronomy. We also ran several activities, such as small physics experiments and virtual planetarium, to introduce the concept of the constellations and optical observations.

\section{Project communication and promotion}
The main way to communicate our project and ideas is gathered on our website, which currently counts more than 3520 unique visitors\footnote{All the statistics in this section concern the first $6$ months of \ts.}. 
The most visited page of the website is the Astrophysics articles followed by Space Physics articles. Our audience frequently visits the section about "Who we are", which describes our team members. This implies that the reader is interested to know who is providing the information and who is behind our project. It is also interesting to note that the mean reading time of our articles is just 2 minutes and 12 seconds, implying that we achieved our goal as \ts!

We are currently advertising our website and overall activities through social media, mainly via Facebook and Instagram. During the last years, social media platforms have gained increasing popularity, becoming one of the main forms of engagement and source of information, especially in the young population. With the aim to attract more and more people towards Astrophysics and scientific thinking, we decided that an active presence in social media is necessary. Currently, we count more than 3000 likes on our Facebook\footnote{\href{https://www.facebook.com/2sciencegr/}{https://www.facebook.com/2sciencegr}} page and more than 750 followers on Instagram\footnote{\href{https://www.instagram.com/2sciencegr}{https://www.instagram.com/2sciencegr}}. Our activity includes daily posts of our articles' highlights, the fun facts, and the answers to the readers' questions. All of our posts lead to our website, where the reader can find the whole article, get information about the authors and have access to extra resources for each topic. 
We additionally include a bi-weekly quiz on Instagram with questions based on our recent articles in order to stimulate learning and involve our audience. This kind of engagement seems to be one of the favorite ones and drives our audience to ask more questions about the discussed topics. We show a summary of our audience demographics in \cref{table}.

\begin{table}[]
    \centering
    \begin{tabular}{c|c|c|c}
         & Webpage & Facebook & Instagram\\
         \hline
    Total users/followers   & 4058 & 3055 & 788 \\
    Total reach & 60.519 & 115.130 & 7217 \\
    Top 3 countries & Greece, USA, Germany & Greece, Germany, UK  & Greece, Germany, UK \\
    \end{tabular}
    \caption{Summary of our audience demographics (estimated as of late June 2021). The total amount of unique users (website), the followers (Facebook, Instagram) and the main location of our followers are listed.}
    \label{table}
\end{table}

\begin{figure}
     \centering
     \includegraphics[width=0.8\textwidth]{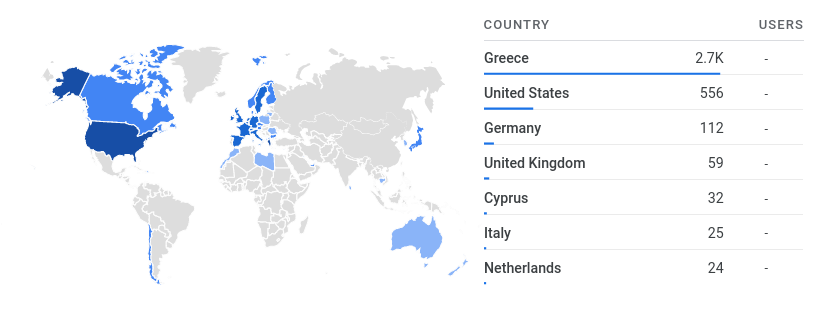}
     \caption{Geographical distribution of the visitors of our website. Credit: Google Analytics}
     \label{fig:map}
 \end{figure}

The age span of our audience peaks at the age group of $25-34$ in both Facebook and Instagram. On all platforms, the gender balance of the audience seems to hold. In \cref{fig:map} we demonstrate the geographical distribution of the visitors of the website. People from more than 40 countries have already visited our website. As expected, the majority of them are located in Greece. However, it is impressive that we have followers that are located all around the globe, from the United States all the way to Australia.

\begin{figure}
    \centering
    \includegraphics[width=0.9\textwidth]{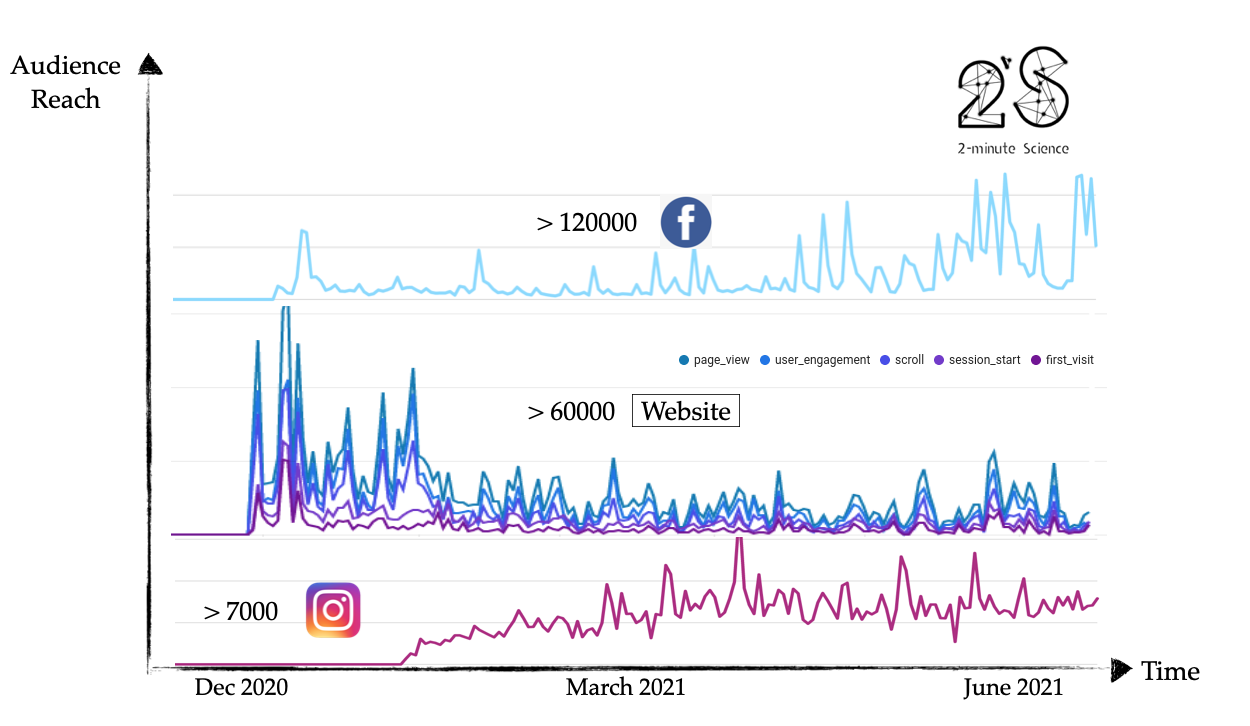}
    \caption{Audience reach from the start of our social media presence. Credit: Google Analytics / Facebook}
    \label{fig:reach}
\end{figure}

We present the reach of our activity in \cref{fig:reach}. Our main promotion method is through our Facebook page. It is also worth mentioning that we did not invest any money in the promotion of our work but the current audience reach is solely from our social media activity. 

Apart from the traditional social media like Facebook and Instagram, we recently expanded on Spotify\footnote{\href{https://cutt.ly/xbXyWHt}{https://open.spotify.com/show/2sciencegr}}. Our main aim is to provide the audio version of all our articles in order to be accessible to the population with vision difficulties. In addition, driven by the rapid pace of life, nowadays people tend to be more eager to immerse in an audio or video format of information rather than spending time reading. 

Last but not least, we are open for collaboration with other scientific projects, such as other podcasts, social media pages (e.g. need more space\footnote{\href{https://www.instagram.com/needmorespacegr/}{https://www.instagram.com/needmorespacegr}}) and other outreach initiatives, like the Activities in High-Energy Astrophysics (AHEAD\footnote{\href{http://ahead.astro.noa.gr}{http://ahead.astro.noa.gr}}), a project lead by scientists at the National Observatory of Athens. A characteristic example includes our invitation in the ``Not a top 10'' podcast\footnote{\href{https://www.instagram.com/notatop10/}{https://www.instagram.com/notatop10}} , where we discussed about black holes and specifically the first portrait of the black hole of M87, about the detection of the first gravitational wave signal and the physics surrounding these objects.

\section{Future plans}

2'science is a very "young" project. However, it is also very ambitious, and there are numerous plans to extend our activities. First of all, our obvious next step is to give the opportunity to more scientists to reach the public. 
After the Covid-19 restrictions are lifted, we are preparing a number of activities and small projects:

- \textbf{In-person activities around Greece.} We plan in-person activities for kids and adults. These range from ``treasure hunts'' with topics related to Astrophysics, Astronomy, and Space Physics, to experiments, such as ``how to cook a comet'' and ``how nebulae behave''. Furthermore, we would like to give talks in different locations around Greece, especially in areas further away from the main cities, where access to observatories, higher education institutes, and other research facilities is limited.   

- \textbf{Interviews with Greek astronomers.} We believe it is essential to have a record of publicly available interviews with established Greek astronomers. The idea is simple: small conversations between astronomers in an informal environment. This would be interesting not only for historical purposes but also, and most importantly, to motivate more people into science. 

- \textbf{Presentations of the history of Greek Astronomy.} Our goal is to create a timeline with important dates and moments of Greek Astronomy. 

- \textbf{Podcast(s) with discussion.} As already mentioned above, our initial podcast episodes correspond to recordings of our articles, allowing for greater accessibility. As our next step, we aim to have small discussions on Astrophysics' "hot" topics. We also want to have round-table conversations on specific topics related to our audience's questions. 

- \textbf{Educational videos.} Optical material - graphics/videos - are proved to be very useful when trying to explain difficult topics. For this reason, we plan to develop dedicated content to supplement our articles or interesting astrophysical phenomena.

- \textbf{Dedicated workshops on specific subjects.} There are many topics that we can cover through dedicated workshops. 
People tend to better comprehend scientific concepts through experiments and hands-on sessions.
Therefore, we plan to organize a series of hands-on workshops of longer duration with topics such as gravitational waves, radio and X-ray Astronomy, where students and adults will have the opportunity to learn about the data analysis procedure,  observational techniques, and how to present scientific results.

- \textbf{Educational activities for students.} We plan to create a database of scientific activities and experiments for students and children. This will be accessible to educators through our website and could be potentially used in schools.

- \textbf{Astronomers' Point of contact.} Usually finding an expert in a specific field, who will be interested in giving a public talk can be difficult. We aim to facilitate the connection between public organisations seeking scientists passionate for science communication, by providing a reference list of potential speakers, based on different fields and regions. 

\section{Summary \& Discussion}

\ts\ is based on our mutual passion for Astrophysics and science communication. As young researchers, we consider the engagement with the public an important task along with our day-to-day scientific work.

Stimulating research and exciting results should be communicated to society. Nevertheless, we believe that science communication should not be in a one-way fashion, but in a more interactive manner where the public can pose its questions and participate in (open) discussion with the experts. In that way, the interested public can further its understanding of the Universe and its physical laws. We believe that this approach helps the readers achieve  physical intuition while at the same time fosters ``critical thinking''. 

The goal of \ts\ is to communicate the aforementioned philosophy together with the most exciting scientific results to the Greek public. We try to accomplish this by a variety of methods: short articles, Q\&As, fun facts, social media, podcasts, and interactive activities. We moreover aim to create a lasting platform that could be utilised in turn by new generations of Astrophysics researchers, interested in science communication. We are more interested in contributing to Greek public science engagement a high-quality and enduring legacy of physics education, rather than building a short-term project. 

Despite our name (\ts), we strongly believe that science cannot be taught within two minutes, and such articles cannot substitute in-detail lectures in schools and universities. Nevertheless, we believe that such an effort can provide extra material to the interested readers and inspire young people to learn more about the Universe.

Among other big questions of the Universe and space technologies, we try to bring relatively young disciplines closer to the public. High-Energy Astrophysics and Astroparticle physics are just two examples of modern scientific fields that are not very popular in the Greek society. Finally, building on different approaches and techniques for science communication, we hope that we can play our part in making science accessible and approachable to all!\\

\small{
\textbf{Acknowledgments}}\\
\footnotesize{
We acknowledge the support of the Hellenic Astronomical Society. We would also like to thank: Nafsika Sioropoulou for creating the logo of \ts\,, Konstantinos Droudakis for making the video presentation, our team members Savvas Raptis, Stefanos Tsiopelas and Eugene Zhuleku for useful comments on the manuscript, and the rest team members Katerina Dima, Dimitra Ligri, Georgia Loukaidou, and all the guest authors who contributed to our project.}


\end{document}